\documentclass[floatfix,aps,pra,reprint]{revtex4-1}
\usepackage{amsmath}
\usepackage{graphicx}
\usepackage{float}
\usepackage[normalem]{ulem}
\usepackage[dvipsnames]{xcolor}
\DeclareMathOperator{\sech}{sech}
\usepackage[hyperindex,breaklinks,hidelinks]{hyperref}

\begin{document}


\title{Theory of Kerr frequency combs in Fabry-Perot resonators}


\author{Daniel C. Cole}
\email[]{daniel.cole@nist.gov}
\affiliation{National Institute of Standards and Technology (NIST), Boulder, CO, 80305 USA}
\affiliation{Department of Physics, University of Colorado, Boulder, CO, 80309 USA}

\author{Alessandra Gatti}
\affiliation{Dipartimento di Scienza e Alta Tecnologia, Universit\`a dell'Insubria, Via Valleggio 11, 22100 Como, Italy}

\author{Scott B. Papp}
\affiliation{National Institute of Standards and Technology (NIST), Boulder, CO, 80305 USA}
\affiliation{Department of Physics, University of Colorado, Boulder, CO, 80309 USA}

\author{Franco Prati}
\affiliation{Dipartimento di Scienza e Alta Tecnologia, Universit\`a dell'Insubria, Via Valleggio 11, 22100 Como, Italy}

\author{Luigi Lugiato}
\email[]{luigi.lugiato@uninsubria.it}
\affiliation{Dipartimento di Scienza e Alta Tecnologia, Universit\`a dell'Insubria, Via Valleggio 11, 22100 Como, Italy}


\date{\today}

\begin{abstract}
We derive a spatiotemporal equation describing nonlinear optical dynamics in Fabry-Perot (FP) cavities containing a Kerr medium. This equation is an extension of the equation that describes dynamics in Kerr-nonlinear ring resonators, referred to as the Lugiato-Lefever equation (LLE) due to its formulation by Lugiato and Lefever in 1987. We use the new equation to study the generation and properties of Kerr frequency combs in FP resonators.
The derivation of the equation starts from the set of Maxwell-Bloch equations that govern the dynamics of the forward and backward propagating envelopes of the electric field coupled to the atomic polarization and population difference variables in a FP cavity. The final equation is formulated in terms of an auxiliary field $\psi(z,t)$ that evolves over a slow time $t$ on the domain $-L \leq z \leq L$ with periodic boundary conditions, where $L$ is the cavity length. We describe how the forward and backward propagating field envelopes can be obtained after solving the equation for $\psi$. This formulation makes the comparison between the FP and ring geometries straightforward. The FP equation includes an additional nonlinear integral term relative to the LLE for the ring cavity, with the effect that the value of the detuning parameter $\alpha$ of the ring LLE is increased by an amount equal to twice the spatial average of $|\psi|^2$. This feature establishes a general connection between the stationary phenomena in the two geometries. 
For the FP-LLE, we discuss the linear stability analysis of the flat stationary solutions, analytic approximations of solitons, Turing patterns, and nonstationary patterns. We note that Turing patterns with different numbers of rolls may exist for the same values of the system parameters. We then discuss some implications of the nonlinear integral term in the FP-LLE for the kind of experiments that have been conducted in Kerr-nonlinear ring resonators.

\end{abstract}


\maketitle

\section{Introduction}
Optical frequency combs have revolutionized the measurement of optical frequencies and enabled a wide array of basic research applications in fields such as time-keeping, cosmology, and astronomy \cite{Jones2000,Udem2002,Diddams2010}. Now, the realization of broadband frequency combs using the whispering gallery modes of high-Q ring microresonators with the Kerr nonlinearity (first described in \cite{DelHaye2007}) promises to bring the capabilities of frequency combs to a new set of applications outside the laboratory. In contrast with mode-locked laser-based frequency combs, microresonator-based Kerr frequency combs arise from the parametric four-wave mixing (FWM) processes activated by the interaction between the driving field and the Kerr medium. The potential of these combs for applications relies on the fact that, under suitable conditions, the newly-generated frequency components can mode-lock to form well-behaved dissipative Kerr-cavity solitons \cite{Herr2014wArxiv,Papp2014,Liang2015,Yi2015,Xue2015,Brasch2016,Cole2017}. These combs can yield natively octave-spanning spectra \cite{Spencer2017,Briles2017}, and they can be regarded as novel multiwavelength sources where all the lines except for the pump laser are created by the gain induced by the FWM processes.

Kerr-microresonator-based frequency combs (microcombs) are anticipated to have a significant impact as a compact, low cost, low power technology. Microcavities can be conveniently pumped with a variety of laser wavelengths, can be embedded on photonics chips, can be integrated in fiber networks, and are compatible with CMOS/metal oxide semiconductors. These properties make microcombs quite promising and have inspired a worldwide effort to develop the technology \cite{Chembo2016}.

The effects of the Kerr nonlinearity in passive, driven optical cavities were analyzed in the 1970s in the field of optical bistability \cite{Gibbs1985,Lugiato1984}.  In this context the possibility that such systems can spontaneously emit cavity modes different from the mode quasi-resonant with the injected driving frequency, then referred to as the multimode instability, was theoretically predicted \cite{Lugiato1984,Bonifacio1978,Bonifacio1979} and experimentally observed \cite{Segard1989} before the concept of a frequency comb was introduced. 
 
As shown e.g. in Refs. \cite{Herr2014wArxiv,Matsko2011,Coen2013a,Coen2013,Chembo2013,Godey2014} the model that is appropriate for the description of comb generation in Kerr resonators and for the exploration and prediction of comb characteristics is an equation formulated thirty years ago by one of us and Lefever  \cite{Lugiato1987,Lugiato1987a} in our investigation of optical bistability. This model is referred to as the Lugiato-Lefever equation (LLE) in the field of microcombs. The equation was originally formulated to provide a paradigm for transverse spatial pattern formation \`{a} la Turing \cite{Turing1952} in nonlinear optical systems, which arises through the simultaneous effects of Kerr nonlinearity and diffraction. The temporal/longitudinal version of the LLE that describes Kerr-comb formation was introduced some years later (\cite{Haelterman1992}, also \cite{Brambilla1992,Castelli2017}) and is characterized by the replacement of  diffraction by group velocity dispersion. It is mathematically equivalent to the transverse LLE in 1D. This equation has been applied both to fiber ring cavities (e.g. \cite{Jang2015,Luo2015}) and to high-Q microresonators (e.g. \cite{Herr2014wArxiv,Matsko2011,Coen2013a,Coen2013,Chembo2013,Godey2014}). The spontaneous formation of traveling spatiotemporal patterns along the ring cavity, described by the LLE, corresponds to the generation of new optical frequencies. It is a remarkable development that the rather idealized physical conditions assumed in the LLE thirty years ago have been perfectly realized by recent progress in the field of photonics, and the LLE and higher order corrections to it have provided a framework in which the vast majority of experimental microcomb results are now understood. In this way the investigations of pattern formation based on the LLE have acquired significant practical importance.

\begin{figure}[]
\includegraphics{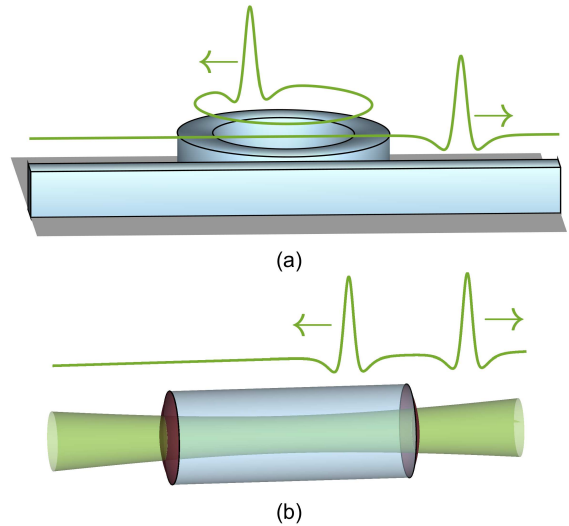} %
\caption{Experimental geometries for microresonator frequency combs. (a) The ring geometry. A pump laser is coupled into the ring resonator through a bus waveguide, and the intracavity intensity envelope (here a soliton pulse) is coupled out once per round trip. (b) The Fabry-Perot geometry. A pump laser is coupled in through a cavity mirror, and the intensity envelope is coupled out at each reflection. Not shown is the background standing wave in the Fabry-Perot cavity, which is made up of forward-propagating and backward-propagating field components. \label{ResGeometry}}
\end{figure}

An interesting variation on microcomb experiments, which to our knowledge has only recently been performed for the first time (\cite{Obrzud2017}, see also \cite{Braje2009}), is the generation of cavity solitons in resonators with the Fabry-Perot (FP) geometry. This is depicted schematically in Fig. \ref{ResGeometry}. In their study, Obrzud and colleagues report on dissipative Kerr-cavity soliton generation in a passive, high-Q Fabry-Perot resonator constructed of standard (anomalous dispersion) single-mode fiber (SMF) with high-reflectivity end-coatings. It is worth mentioning that this work also makes use of a pulsed pump laser (see also e.g. Ref. \cite{Lobanov2016}).

Practical differences with the ring geometry make microcomb generation in FP cavities appealing. In particular, the FP geometry offers different methods for tailoring the cavity dispersion, which dictates the bandwidth and temporal duration of cavity solitons. Engineering of the core-cladding index contrast is analogous to engineering of the geometrical dispersion in ring resonators \cite{Yang2016}, while there exists for FP cavities the additional opportunity to employ chirped mirror end-coatings. Further, in the particular case of an FP cavity constructed of SMF the cavity naturally has a single transverse mode family, which avoids the practical difficulties of comb generation in a resonator populated with higher-order mode families, each with its own free spectral range.

In this article we provide a theoretical treatment of the nonlinear dynamics in a passive, driven FP cavity containing a Kerr medium as they apply to frequency comb formation. A brief treatment within the formalism of coupled-mode equations was provided by Obrzud \textit{et al.} in Ref. \cite{Obrzud2017}. Here, we derive and explore the properties of a complementary spatiotemporal equation analogous to the LLE for the FP geometry. This FP-LLE differs from the ring LLE in the existence of an additional nonlinear term that represents phase modulation by the average of the intracavity intensity. The effects of this term are similar to the effects of the thermal shift of the cavity resonance frequency present in all microresonator experiments \cite{Herr2014wArxiv,Carmon2004}, but the additional nonlinear term acts on the timescale of the Kerr nonlinearity, which is effectively instantaneous. The new term connects stationary patterns in the Fabry-Perot resonator to stationary patterns in the ring resonator with a shifted detuning parameter, and it imparts a dispersion-dependence to the region of parameter space over which solitons can exist. We explore in particular the generation of single solitons through laser frequency sweeps, which is commonplace for ring resonators, and find that the additional nonlinear term may present new challenges that can be alleviated by using high pump power or a pulsed pump laser.

In Sec. \ref{sec2} we derive a set of two coupled equations for the forward- and backward-propagating electric field components in a Kerr-nonlinear FP resonator. In Sec. \ref{sec3} we derive the generalization of the LLE to the FP geometry from these coupled equations. We show how to reverse the procedure and obtain the two counter-propagating fields from the solution of the FP-LLE. 

Sec. \ref{sec4} is devoted to the homogeneous stationary solutions of the FP-LLE and to their linear stability analysis; in both cases we compare with the ring cavity case. In addition, we demonstrate a general connection that links the stationary patterns of the FP-LLE with those of the ring LLE. 

In Sec. \ref{sec5} we focus on the soliton solutions of the LLE. We review a well-known analytical expression obtained in the ring case and extend it to the FP case. 

In Sec. \ref{sec6} we explore Turing patterns under the FP-LLE, and in particular their multi-stability. 

In Sec. \ref{sec7} we discuss nonstationary phenomena in the FP-LLE, including spatiotemporal chaos and oscillating breather solitons. 

In Sec. \ref{sec8} we discuss implications for experiments of the differences between the dynamics under the FP-LLE and dynamics under the ring LLE, which have been well explored experimentally. 

Finally, in Sec. \ref{sec9} we conclude with some general remarks.

\section{Derivation of coupled equations for the forward- and backward-propagating envelopes\label{sec2}}

To derive the temporal/longitudinal version of the LLE for the ring cavity case, one must start from an equation (Eq. (7.19) of Ref. \cite{Lugiato2015}) that governs the propagation of the slowly varying envelope of the electric field in a Kerr medium in the presence of second-order chromatic dispersion.  To generalize this procedure to the FP case, a set of coupled equations for the forward- and backward-propagating field envelopes in the cavity must be derived. The main problem is to correctly formulate the cubic terms that describe the Kerr nonlinearity this two-field configuration. To do this, we generalize to the two-field case the derivation of the temporal/spatial LLE for the ring cavity from the Maxwell-Bloch equations (MBE), which describe the interaction of the field envelope with a two-level medium, given in Refs. \cite{Brambilla1992,Castelli2017}. 

We start from a set of equations (Eqs. (14.61-64) of Ref. \cite{Lugiato2015}) that provide a convenient generalization of the MBE to the FP case and include the high-Q limit.  These equations represent the analogue of the MBE for the ring cavity, starting from which Ref. \cite{Castelli2017} shows the derivation of the temporal/longitudinal LLE. Obtaining these equations from the MBE is not trivial, and is described in detail in Ref. \cite{Castelli2017}. Briefly, the process involves: 1. Applying a slowly-varying envelope approximation to the MBE for two-level atoms, 2. Enforcing Fabry-Perot boundary conditions, 3.~Taking the high-Q limit, and 4. Distinguishing between the widely-separated scales of the optical wavelength and the cavity length. The resulting equations read:

\begin{multline}
\frac{\partial\tilde{F}_F(z,t)}{\partial t}+\tilde{c}\frac{\partial\tilde{F}_F(z,t)}{\partial z}=
-\kappa\left[\vphantom{\int_{-\pi}^{\pi}}(1+i\alpha)\tilde{F}_F(z,t)-\tilde{F}\right.\\\left.+\frac{C}{\pi}\int_{-\pi}^{\pi}d\phi\, e^{-i\phi}\tilde{P}(z,\phi,t)\right]\label{MBE1},
\end{multline}
\begin{multline}
\frac{\partial\tilde{F}_B(z,t)}{\partial t}-\tilde{c}\frac{\partial\tilde{F}_B(z,t)}{\partial z}=
-\kappa\left[\vphantom{\int_{-\pi}^{\pi}}(1+i\alpha)\tilde{F}_B(z,t)-\tilde{F}\right.\\\left.+\frac{C}{\pi}\int_{-\pi}^{\pi}d\phi\, e^{i\phi}\tilde{P}(z,\phi,t)\right]\label{MBE2},
\end{multline}
\begin{multline}
\gamma_\perp^{-1}\frac{\partial\tilde{P}(z,\phi,t)}{\partial t}=\left[\left(\tilde{F}_F(z,t)e^{i\phi}+\tilde{F}_B(z,t)e^{-i\phi}\right)\times\right. \\
\left.\vphantom{\tilde{F}_F(z,t)e^{i\phi}}D(z,\phi,t)\right]-(1+i\Delta)\tilde{P}(z,\phi,t)\label{MBE3},
\end{multline}
\begin{multline}
\gamma_\parallel^{-1}\frac{\partial D(z,\phi,t)}{\partial t}=-\frac{1}{2}\left[\left\{\left(\tilde{F}_F(z,t)e^{i\phi}+\tilde{F}_B(z,t)e^{-i\phi}\right)\right.\right.\\
\times\left.\left.\tilde{P}(z,\phi,t)\right\}+c.c.\vphantom{\tilde{P}e^{i\phi}}\right]-D(z,\phi,t)+1\label{MBE4},
\end{multline}
where $\tilde{F}_F(z,t)$, $\tilde{F}_B(z,t)$, and $\tilde{F}$ denote the normalized envelopes of the forward and backward propagating fields and of the input field, respectively, and $\tilde{P}(z,\phi,t)$ and $D(z,\phi,t)$ indicate the normalized atomic polarization and population difference of the two-level atoms, respectively. The speed of light in the medium is $\tilde{c}=c/n$, with $c$ the speed of light in vacuum and $n$ the background refractive index. The transverse and longitudinal atomic relaxation rates are $\gamma_\perp$ and $\gamma_\parallel$, respectively, and the atomic detuning parameter is $\Delta=(\omega_a-\omega_o)/\gamma_\perp$, where $\omega_o$ is the frequency of the driving field and $\omega_a$ is the Bohr transition frequency of the two-level atoms; $C$ is the bistability parameter \cite{Lugiato1984,Lugiato2015}. Time is indicated by $t$, while there are two distinct spatial variables: the slow spatial variable $z$, which varies on the scale of the cavity length, and the fast spatial variable $\phi=\omega_o z/\tilde{c}$, which varies from $-\pi$ to $+\pi$ and is related to the wavelength scale. The cavity damping rate is defined as $\kappa=\tilde{c}T/2L$, where $L$ is the cavity length. The cavity detuning is given by $\alpha=(\omega_c-\omega_o)/\kappa$ (in many of the references relevant to this derivation this quantity is represented by $\theta$, which here we reserve for another purpose), with $\omega_c$ being the cavity frequency closest to $\omega_o$. As usual $c.c.$ means complex conjugate.

The electric field $ E (z,t) $, assumed linearly polarized for simplicity, is expressed as
\begin{multline}
E(z,t)=\frac{1}{2}\frac{\hbar\sqrt{\gamma_\perp\gamma_\parallel}}{d}(\tilde{F}_F(z,t)e^{-i\omega_o(t-\frac{z}{\tilde{c}})}\\
+\tilde{F}_B(z,t)e^{-i\omega_o(t+\frac{z}{\tilde{c}})}+c.c.)\label{Efield},
\end{multline}
where $d$ is the modulus of the atomic dipole moment and $\hbar$ is Planck's constant. The two exponentials that appear in Eq. (\ref{Efield}) can be rewritten as $e^{-i(\omega_o t-\phi)}$ and $e^{-i(\omega_o t+\phi)}$. The electric field injected into the cavity is given by   
\begin{equation}
E_I=\frac{1}{2}\frac{\hbar\sqrt{T\gamma_\perp\gamma_\parallel}}{d}(\tilde{F}e^{-i\omega_o t}+c.c.),
\end{equation} 
where $T$ is the transmissivity coefficient of the cavity mirrors. The forward- and backward-propagating fields obey the boundary conditions
\begin{equation}\label{BCs}
\tilde{F}_F(0,t)=\tilde{F}_B(0,t), \quad \tilde{F}_F(L,t)=\tilde{F}_B(L,t).
\end{equation}

To derive two coupled self-contained equations for the counterpropagating fields, we follow the same steps described in the Appendix of Ref. \cite{Castelli2017}, starting from Eqs.~(\ref{MBE1}-\ref{MBE4}) instead of the MBE for the ring cavity. 
The derivation is not described here in detail, but introduces the following assumptions:
\begin{itemize}
\item The dispersive limit $|\Delta|\gg 1$, i.e. the central frequency $\omega_o$ is far off-resonance from the atomic line, and $\Delta<0$.
\item The bandwidth of the fields is small in comparison with the detuning of the central frequency.
\item We assume that $|\tilde{F}_F/\Delta|\ll 1$ and $|\tilde{F}_B/\Delta|\ll 1$, which allows truncation of the power expansion of the atomic polarization in terms of the counterpropagating fields after the first terms.
\item The radiative limit $\gamma_\parallel=2\gamma_\perp$. 
\end{itemize}

If we define normalized fields as
\begin{equation}
F=\sqrt{\frac{2C}{|\Delta|^3}}\tilde{F},\label{normfields}
\end{equation}
where the same scaling factor also relates $\tilde{F}_F$ to $F_F$ and $\tilde{F}_B$ to $F_B$, and
\begin{equation}
\alpha_o=\alpha-\frac{2C}{\Delta},
\end{equation}
\begin{equation}
\bar{a}=\frac{2Cv_g^2}{\Delta^3\gamma_\perp^2}<0,
\end{equation}
we arrive at the two coupled equations
\begin{multline}
\frac{\partial F_F(z,t)}{\partial t}+v_g\frac{\partial F_F(z,t)}{\partial z}=
-\kappa\left[\vphantom{\frac{\partial^2}{\partial}}(1+i\alpha_o)F_F-F\right.\\\left.-i(|F_F|^2+2|F_B|^2)F_F+i\bar{a}\frac{\partial^2F_F}{\partial z^2}\right]\label{Catomic1},
\end{multline}
\begin{multline}
\frac{\partial F_B(z,t)}{\partial t}-v_g\frac{\partial F_B(z,t)}{\partial z}=
-\kappa\left[\vphantom{\frac{\partial^2}{\partial}}(1+i\alpha_o)F_B-F\right.\\\left.-i(|F_B|^2+2|F_F|^2)F_B+i\bar{a}\frac{\partial^2F_B}{\partial z^2}\right]\label{Catomic2}.
\end{multline}
Here the second derivative terms describe anomalous dispersion as explained in Ref. \cite{Castelli2017}. The first nonlinear term describes self-phase modulation and the second describes cross-phase modulation, while the components of the term $(1+i\alpha_o)$ describe loss and detuning, respectively. The group velocity $v_g$ is given by
\begin{equation}
v_g=\tilde{c}\left(1+\frac{2C\kappa}{\Delta^2\gamma_\perp}\right)^{-1}\approx\tilde{c}\left(1-\frac{2C\kappa}{\Delta^2\gamma_\perp}\right)\approx\tilde{c}.
\end{equation}

We now seek the same equations in the general case of a Kerr medium with chromatic dispersion of second order. By following the analysis in Sec. 28.2.1 of Ref.~\cite{Lugiato2015} we find that if the electric field and the input field are given by
\begin{multline}
E(z,t)=\frac{1}{2}\sqrt{\frac{4}{3}\frac{(1-R)}{L}\frac{cn}{\omega_o}\frac{1}{\chi^{(3)}}}\times\\
\left(F_Fe^{-i\omega_o(t-\frac{z}{v_g})}+F_Be^{-i\omega_o(t+\frac{z}{v_g})}+c.c.\right), \label{EfieldKerr}
\end{multline}
\begin{equation}
E_I=\frac{1}{2}\sqrt{\frac{4}{3}\frac{(1-R)^3/T}{L}\frac{cn}{\omega_o}\frac{1}{\chi^{(3)}}}
(Fe^{-i\omega_o t}+c.c.),
\end{equation}
where $\chi^{(3)}$ is the third-order nonlinear susceptibility and $R$ is the reflectivity of the cavity mirrors, then the coupled equations in this case coincide with Eqs. (\ref{Catomic1}) and (\ref{Catomic2}) provided that $\kappa$ is defined as $\kappa=v_g (1-R)/2L$, $\alpha_o$ is replaced with $\alpha$, and $\bar{a}$ is defined as 
\begin{equation}
\bar{a}=k^{\prime\prime}Lv_g^2/T, 
\end{equation}
where
\begin{equation}
k^{\prime\prime}=\left.\frac{\partial^2k(\omega)}{\partial\omega^2}\right|_{\omega=\omega_o},
\end{equation}
with $k(\omega)$ being the dispersion law; $k^{\prime\prime}>0$ in the case of normal dispersion and $k^{\prime\prime}<0$ in the case of anomalous dispersion. Eqs. (\ref{Catomic1}) and (\ref{Catomic2}) include the correct form for the Kerr nonlinear terms in the case of two counterpropagating fields. The above treatment allows for mirror losses in the Fabry-Perot cavity such that $T<1-R$ is possible (but not required).

A final comment in this section is the following. In the case of unidirectional propagation (i.e. for $F_B (z,t)=0$) Eq. (\ref{Catomic1}) is equivalent to the temporal/longitudinal version of the LLE; as a matter of fact, by simply transforming Eq. (\ref{Catomic1}) from the variables $(z,t)$ to the variables $(t,\bar{t}=t-z/v_g)$ one obtains the LLE formulated in Ref. \cite{Haelterman1992}. The same trick is not possible in the case of Eqs. (\ref{Catomic1}) and (\ref{Catomic2}) because they involve two distinct retarded times, one for forward propagation and the other for backward propagation. This implies that one must solve numerically the two equations calculating the forward propagation in the cavity and then the backward propagation and so on, so that an exceedingly high number of roundtrips are necessary to reach the long time scale, on the order of the inverse of $\kappa$, which governs the relaxation of the system to a steady state. Such a calculation is not practical in the high-$Q$ limit.

\section{The LLE for Fabry-Perot Resonators\label{sec3}}
\subsection{Derivation of a single envelope equation from the modal expansion}
To unite Eqs. (\ref{Catomic1}) and (\ref{Catomic2}) into a single spatiotemporal equation describing dynamics in the cavity, we next introduce a modal expansion for the fields $F_F$ and $F_B$ in terms of the modal amplitudes $\bar{f}_\mu$ (similar to that found in (A.20) of Ref. \cite{Castelli2017}):
\begin{equation}
F_F(z,t)=\sum_{\mu=-\infty}^\infty \bar{f}_\mu(t)e^{i\frac{\alpha_\mu}{v_g}z},\label{FFexp}
\end{equation}
\begin{equation}
F_B(z,t)=\sum_{\mu=-\infty}^\infty \bar{f}_\mu(t)e^{-i\frac{\alpha_\mu}{v_g}z},\label{FBexp}
\end{equation}
where $\alpha_\mu$ is defined as:
\begin{equation}
\alpha_\mu=\pi \mu v_g/L.\label{alphan}
\end{equation}
Using the expansions in Eqs. (\ref{FFexp}) and (\ref{FBexp}) one can extend the functions $F_F(z,t)$ and $F_B(z,t)$ to the interval $-L\leq z\leq L$. This amounts to defining $F_F(z,t)$ and $F_B(z,t)$ for $-L\leq z \leq 0$ as
\begin{equation}
F_F(z,t)=F_B(-z,t), \quad F_B(z,t)=F_F(-z,t),\label{negzeqs}
\end{equation}
and by using Eqs. (\ref{BCs}), (\ref{normfields}), and (\ref{negzeqs}), one sees that $F_F(z,t)$ and $F_B(z,t)$ obey periodic boundary conditions over the interval $-L \leq z \leq L$. Thus, the modal amplitudes $\bar{f}_\mu$ can be obtained as:
\begin{align}
\bar{f}_\mu(t)&=\frac{1}{2L}\int_{-L}^{+L}dz\,e^{-i\frac{\alpha_\mu}{v_g}z}F_F(z,t),\label{FFint}\\
&=\frac{1}{2L}\int_{-L}^{+L}dz\,e^{i\frac{\alpha_\mu}{v_g}z}F_B(z,t).\label{FBint}
\end{align}
Next, we insert Eqs. (\ref{FFexp}) and (\ref{FBexp}) into Eq. (\ref{Catomic1}) with $\alpha_o$ replaced by $\alpha$ and, by using Eqs. (\ref{FFint}) and (\ref{FBint}), we obtain the following set of ordinary differential equations:
\begin{multline}
\frac{d\bar{f}_\mu}{dt}=-i\alpha_\mu\bar{f}_\mu-\kappa\left[\vphantom{\sum_{\mu^\prime \mu^{\prime\prime}}}(1+i\alpha)\bar{f}_\mu-F\delta_{\mu,0}-ia_\mu\bar{f}_\mu\right.\\\left.-i\sum_{\mu^\prime \mu^{\prime\prime}}\bar{f}_{\mu^\prime}\bar{f}_{\mu^{\prime\prime}}^*(\bar{f}_{\mu-\mu^\prime+\mu^{\prime\prime}}+2\bar{f}_{-\mu+\mu^\prime+\mu^{\prime\prime}})\right],
\end{multline}
where
\begin{equation}
a_\mu=\bar{a}\left(\frac{\pi \mu}{L}\right)^2.\label{an}
\end{equation}
The same set of equations can also be obtained by inserting Eqs. (\ref{FFexp}) and (\ref{FBexp}) into Eq. (\ref{Catomic2}). If we now define
\begin{equation}
\bar{f}_\mu(t)=f_\mu(t)e^{-i\alpha_\mu t},\label{fbarnt}
\end{equation}
so that Eqs. (\ref{FFexp}) and (\ref{FBexp}) read
\begin{equation}
F_F(z,t)=\sum_{\mu=-\infty}^\infty f_\mu(t)e^{-i\alpha_\mu(t-\frac{z}{v_g})},\label{FFexptime}
\end{equation}
\begin{equation}
F_B(z,t)=\sum_{\mu=-\infty}^\infty f_\mu(t)e^{-i\alpha_\mu(t+\frac{z}{v_g})},\label{FBexptime}
\end{equation}
the set of differential equations becomes
\begin{multline}
\frac{df_\mu}{dt}=-\kappa\left[\vphantom{2f_{-\mu+\mu^\prime+\mu^{\prime\prime}}e^{2i(\alpha_\mu-\alpha_{\mu^{\prime}})t}}(1+i\alpha)f_\mu-F\delta_{\mu,0}-ia_\mu f_\mu\right.\\
-i\sum_{\mu^\prime \mu^{\prime\prime}}f_{\mu^\prime}f_{\mu^{\prime\prime}}^*\left(\vphantom{2f_{-\mu+\mu^\prime+\mu^{\prime\prime}}e^{2i(\alpha_\mu-\alpha_{\mu^{\prime}})t}}f_{\mu-\mu^\prime+\mu^{\prime\prime}}\right. \\
	\left.\left.+2f_{-\mu+\mu^\prime+\mu^{\prime\prime}}e^{2i(\alpha_\mu-\alpha_{\mu^{\prime}})t}\right)\right]. \label{fnDecomp}
\end{multline}
Equation (\ref{fbarnt}) decomposes $\bar{f}_\mu(t)$ into the product of two functions that vary on two distinct time scales. The exponential varies on the scale of the roundtrip time $T_{RT}=2L/v_g$, while $f_\mu(t)$ varies on the scale of the cavity decay time $\kappa^{-1}=2L/v_gT$. For a high-$Q$ cavity $T$ is much smaller than $1$, so that the two scales are widely separated. If one averages the terms of Eq. (\ref{fnDecomp}) over a time interval much longer than the cavity roundtrip time but much shorter than the cavity decay time, all terms of Eq. (\ref{fnDecomp}) remain unchanged except the last, which vanishes in the average for $\mu^\prime\neq \mu$. Therefore in the last term we set $\mu^\prime=\mu$, obtaining
\begin{multline}
\frac{df_\mu}{dt}=-\kappa\left[(1+i\alpha)f_\mu-F\delta_{\mu,0}-ia_\mu f_\mu\vphantom{\sum_{\mu^\prime \mu^{\prime\prime}}}\right.\\
\left.-i\sum_{\mu^\prime \mu^{\prime\prime}}f_{\mu^\prime}f_{\mu^{\prime\prime}}^*f_{\mu-\mu^\prime+\mu^{\prime\prime}}-2if_\mu\sum_{\mu^\prime}f_{\mu^\prime}^*f_{\mu^\prime}\right], \label{fnDecompAvg}
\end{multline}
in agreement with the coupled-mode equations presented in the Supplement of Ref. \cite{Obrzud2017}. 

Basically, to obtain Eq. (\ref{fnDecompAvg}) we neglect the terms that do not conserve energy. The two nonlinear terms in Eq. (\ref{fnDecompAvg}) arise from the two nonlinear terms in Eq. (\ref{Catomic1}), respectively. Therefore, the second nonlinear term represents the difference between the FP cavity and the ring cavity. The first nonlinear term describes processes of self-phase-modulation, cross phase-modulation, and four-wave mixing among modes. The second nonlinear term corrects the coefficients of self-phase-modulation and cross-phase-modulation.

Finally, we define 
\begin{equation}
\psi(z,t)=\sum_{\mu=-\infty}^{+\infty}f_\mu(t)e^{i\frac{\alpha_\mu}{v_g}z},\label{psisum}
\end{equation}
and we obtain from this equation the following partial differential equation for $\psi(z,t)$:
\begin{multline}
\frac{\partial\psi}{\partial t}=-\kappa\left[\vphantom{\int_{-L}^{+L}}(1+i\alpha)\psi - F +i\bar{a}\frac{\partial^2\psi}{\partial z^2}-i|\psi|^2\psi\right.\\
\left.-2i\psi\frac{1}{2L}\int_{-L}^{+L}dz\,|\psi|^2\right]. \label{psiPDE}
\end{multline}
\subsection{The normalized LLE for Fabry-Perot resonators; Connection to experimental parameters}
We now pass to normalized temporal and spatial variables $\tau=\kappa t$ and $\theta=z\cdot\pi/L$, so that when $z$ varies from $-L$ to $L$, $\theta$ varies from $-\pi$ to $\pi$, and we obtain from Eq.~(\ref{psiPDE}) the LLE for the Fabry-Perot cavity as we discuss it throughout the remainder of the paper:
\begin{equation}
\frac{\partial\psi}{\partial\tau}=-(1+i\alpha)\psi+i|\psi|^2\psi-i\frac{\beta}{2}\frac{\partial^2\psi}{\partial\theta^2}+2i\psi\left<|\psi|^2\right>+F.\label{FPLLE}
\end{equation}

Here, $<g>$ denotes the spatial average over the domain: $<g>=\frac{1}{2\pi}\int_{-\pi}^{\pi}d\theta\,g(\theta)$. We have defined
\begin{equation}
\beta=\frac{2\pi^2}{L^2}\bar{a}.\label{beta}
\end{equation}
If we drop the additional nonlinear integral term, Eq. (\ref{FPLLE}) reduces to the temporal/longitudinal LLE in the notations of, e.g., Refs. \cite{Chembo2013} and \cite{Godey2014}. Hence the complete Eq. (\ref{FPLLE}) constitutes the LLE for a Fabry-Perot cavity, with the additional nonlinear integral term $2i\psi\left<|\psi|^2\right>$ representing phase modulation by twice the average intracavity intensity. We stress that we use the spatial variable $\theta$ to make the comparison with the ring cavity case straightforward. 

The FP-LLE is formulated in terms of normalized parameters: $\alpha$ represents the detuning of the pump laser from the nearest cavity resonance, $F^2$ represents the input power, and $\beta$ represents the dispersion. The relationship of these quantities to the experimental parameters is:
\begin{equation}
\alpha=\frac{\omega_c-\omega_o}{\kappa}=-\frac{2(\omega_o-\omega_c)}{\Delta\omega_o},
\end{equation}
\begin{equation}
\beta=\left.-\frac{2D_2}{\Delta\omega_o}=-\frac{2}{\Delta\omega_o}\frac{\partial^2\omega_\mu}{\partial \mu^2}\right|_{\mu=0}, \label{betaLLE}
\end{equation}
\begin{equation}
F^2=\frac{8g_o\Delta\omega_{ext}}{\Delta\omega_o^3}\frac{A_{eff}}{A_{in}}\frac{n_o}{n_{ext}}\frac{P}{\hbar\omega_o}.
\end{equation}
In the above, $\omega_\mu$ represents the set of resonance frequencies of the cavity including the effects of dispersion, with $\mu=0$ indexing the pumped mode (see e.g. Ref.~\cite{Brasch2016}). The full-width-at-half-maximum cavity linewidth is twice the damping rate, $\Delta\omega_o=2\kappa=(1-R)c/n_gL$, and $\Delta\omega_{ext}=cT/2n_gL$ is the coupling rate, with $n_g=c/v_g$ the group index. The quantities $A_{in}$ and $A_{eff}$ represent the mode's effective area $\pi w_{in}^2$ (for a Gaussian mode of radius $w$) at the input mirror and the same averaged over the cavity of length $L$, $\frac{\pi}{L}\int dz\, w(z)^2$, respectively. Further, $g_o=n_2\hbar\omega_o^2 D_1/(2\pi n_g A_{eff})$ is the nonlinear gain parameter, where $D_1=\left.\frac{\partial\omega_\mu}{\partial \mu}\right|_{\mu=0}$ is the cavity free-spectral range in angular frequency (here and in Eq.~(\ref{betaLLE}) $\mu$ is treated as a continuous variable). The nonlinear index $n_2$ is related to the third-order susceptibility via $\chi^{(3)}=(4/3)n_o^2\epsilon_ocn_2$, where $n_o$ is the refractive index of the nonlinear medium.  The power $P=\eta P_{inc}$ denotes the mode-matched power, with mode-matching factor $\eta$ and power $P_{inc}$ incident on the input mirror, and $n_{ext}$ is the refractive index of the medium external to the cavity. 

\begin{figure}[]
	\includegraphics{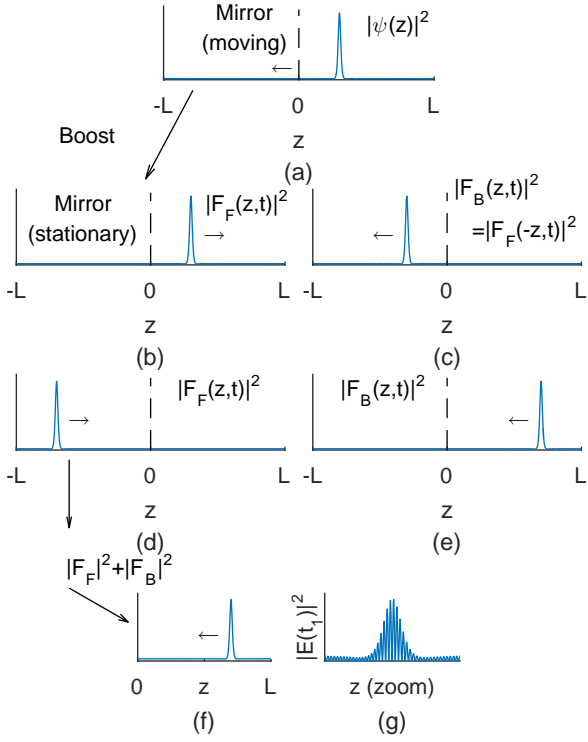} %
	\caption{The relationship between the FP-LLE quantity $\psi$ and the electric field. (a) Soliton-shaped stationary solution of the FP-LLE, Eq. (\ref{FPLLE}), shown in a frame moving in the direction of the forward field (the mirrors move in this frame) so that the field $\psi$ is stationary. (b) Forward-propagating field $F_F$ in the laboratory frame for $t=0$ (see Eq. (\ref{FFexptime})). (c) Backward-propagating field $F_B (z,t)=F_F (-z,t)$ in the laboratory frame for $t=0$ (see Eq. (\ref{FBexptime})). The fields $F_F$ (z,t) and $F_B (z,t)$ obey periodic boundary conditions in the interval $-L\leq z\leq L$. (d,~e) Fields $F_F$ and $F_B$ after propagation for half a round-trip time. The soliton in the forward-propagating field has entered the unphysical region $-L\leq z\leq 0$, and the soliton in the backward-propagating field appears in the physical region $0\leq z\leq L$. (f) The quantity $|F_F|^2+|F_B|^2$ corresponding to panels (d) and (e), proportional to the intensity averaged over fast temporal and spatial oscillations associated with the optical frequency. (g) The physical intensity $|E|^2$, calculated from Eq. (\ref{EfieldKerr}) and shown here at a particular time $t_1$. Cavity parameters have been chosen to facilitate depiction of the standing wave and the soliton on the same scale. \label{Fig1}}
\end{figure}

Fig. \ref{Fig1} schematically depicts the relationship between the FP-LLE quantity $\psi$, the forward- and backward-propagating fields $F_F$ and $F_B$, and the total electric field $E$. The forward and backward fields can be obtained from $\psi(\theta,\tau)$ in the following way: One solves Eq. (\ref{FPLLE}) with periodic boundary condition in the interval $-\pi\leq \theta\leq +\pi$. Then one passes to the original variables $z$, $t$ to obtain $\psi(z,t)$. Next one calculates the coefficients $f_\mu(t)$ using Eq. (\ref{psisum}) as
\begin{equation}
f_\mu(t)=\frac{1}{2L}\int_{-L}^{+L}dz\,e^{-i\frac{\alpha_\mu}{v_g}z}\psi(z,t).
\end{equation}
Finally one obtains $F_F (z,t)$ and $F_B (z,t)$ by utilizing Eqs. (\ref{FFexptime}) and (\ref{FBexptime}). Fig. \ref{Fig1} depicts the behavior of a soliton in the FP cavity as described by Eq. (\ref{FPLLE}). The soliton bounces back and forth in the physical region $0\leq z\leq L$ of the domain, with a period equal to the cavity roundtrip time.

We conclude this section with two remarks. The first starts from the final comment in the previous section, that the problem with Eqs. (\ref{Catomic1}) and (\ref{Catomic2}) lies in the presence of two retarded times. This problem has been solved by introducing the field $\psi(z,t)$ defined by Eq. (\ref{psisum}), which has made the formulation of the LLE for FP cavity straightforward. We note that $\psi(z,t)$ is precisely the field in terms of which the temporal/longitudinal LLE for a ring cavity was formulated in Ref. \cite{Brambilla1992} (where $\psi(z,t)$ is indicated by $X(z,t)$).

The second remark is that the extension of Eqs. (\ref{FFexp}) and (\ref{FBexp}) in the interval $-L \leq z\leq L$ has allowed us to use periodic boundary conditions and therefore traveling waves, which make calculations straightforward. Of course, one can also use the FP boundary conditions in the original interval $0\leq z\leq L$ with standing waves, but one arrives at the same results after calculations that are several times longer.

\section{Flat stationary solutions and their stability; a general connection between stationary patterns in Fabry-Perot and ring resonators\label{sec4}}
\subsection{Flat stationary solutions}
By using Eqs. (\ref{alphan}) and (\ref{psisum}) and the definitions of the normalized variables $\tau$ and $\theta$ we can write
\begin{equation}
\psi(\theta,\tau)=\sum_{\mu=-\infty}^{\infty}f_\mu(\tau)e^{i\mu\theta},
\end{equation}
and the quantities $|f_\mu(\tau)|^2$ constitute the spectrum of the field.

Let us first consider the flat (i.e. homogeneous) stationary solutions of the FP-LLE, Eq. (\ref{FPLLE}); these are the solutions that have no spatial dependence, and are obtained by setting all derivatives to zero. This leads to the stationary equation
\begin{equation}
F=\left[1+i(\alpha-3\rho)\right]\psi_s, \label{statF}
\end{equation}
where $\psi_s$ denotes a flat stationary solution and $\rho=|\psi_s|^2$, so that, assuming that $F$ is real and positive for definiteness, 
\begin{equation}
F^2=\left[1+(\alpha-3\rho)^2\right]\rho. \label{statF2}
\end{equation}

\begin{figure}[]
	\includegraphics{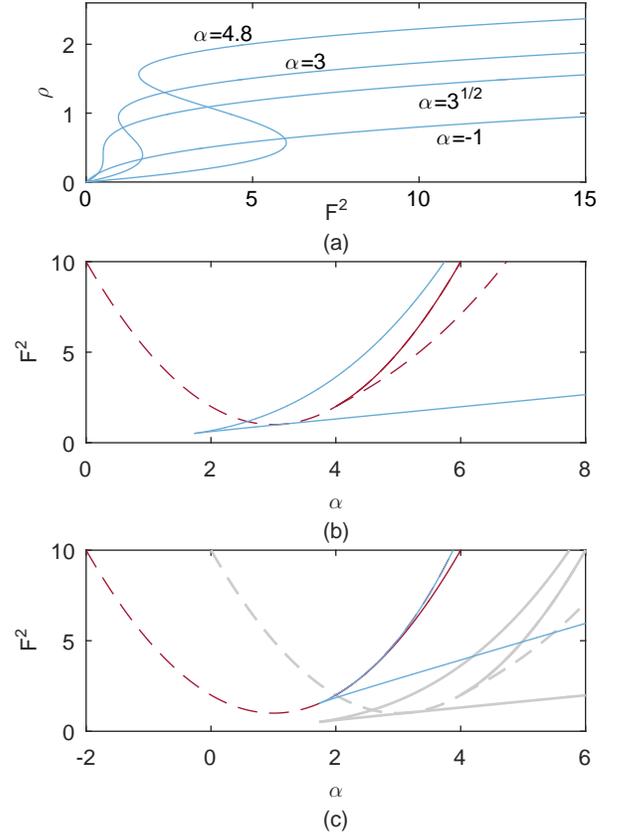} %
	\caption{Analytical curves depicting behavior of the FP-LLE. (a) Stationary curves $\rho$ of normalized transmitted intensity as a function of normalized input intensity $F^2$ for the indicated values of the cavity detuning parameter $\alpha$. (b) Important curves in the $\alpha-F^2$ plane, as discussed in the text. Shown in dashed red is the line obtained from Eq. (\ref{statF2}) by setting $\rho=\rho_{inst}(\alpha)$, where $\rho_{inst}(\alpha)=1$ for $\alpha\leq 4$ and for $\alpha>4$ is given by the solution of the equation $\mu_-(\rho)=0$ with respect to $\rho$, where $\mu_-$ is defined by Eq. (\ref{npm}). The solid red curve is the continuation of the curve obtained from Eq. (\ref{statF2}) by setting $\rho=1$ for $\alpha>4$. Multiple real values of $\rho$ exist in the region bounded by the blue curves, which trace out the local extrema values of the curves $F^2(\rho)$ as a function of $\alpha$. (c)~Analogous curves in the ring cavity. For direct comparison we also plot the FP-cavity curves in light grey. \label{Fig2}}
\end{figure}

By solving this equation for $\rho$ one obtains the function $\rho(\alpha,F^2)$. Considering this function for fixed values of $\alpha$, we obtain the stationary curve of $\rho$ as a function of $F^2$. This curve is single-valued for $\alpha<\sqrt{3}$ and S-shaped for $\alpha>\sqrt{3}$ (as for the ring cavity); in the latter case it displays one stationary solution if $\rho<\rho_-(\alpha)$ or $\rho>\rho_+(\alpha)$, and three stationary solutions for $\rho$ in the interval $(\rho_-(\alpha),\rho_+(\alpha))$. Here, $\rho_\pm$ are defined as the points at which the derivative $\partial F^2/\partial \rho$ vanishes:
\begin{equation}
\rho_\pm=\frac{2\alpha\pm\sqrt{\alpha^2-3}}{9}. \label{derF2zero}
\end{equation}
These results are summarized in Figure \ref{Fig2}: Fig. \ref{Fig2}a shows the stationary curve $\rho$ for several values of $\alpha$, and Fig. \ref{Fig2}b exhibits, in particular, the plots of  $F_-^2 (\alpha)=F^2 (\rho_+ (\alpha))$ and $F_+^2 (\alpha)=F^2 (\rho_- (\alpha))$ obtained from Eqs. (\ref{statF2}) and (\ref{derF2zero}) (see Ref. \cite{Godey2014} for the ring cavity).

In the ring cavity case the factor 3 in Eqs. (\ref{statF}) and (\ref{statF2}) is replaced with $1$, and in Eq. (\ref{derF2zero}) the denominator is $3$ instead of $9$.

\subsection{Linear stability analysis of the flat stationary solutions}
In correspondence with a flat stationary solution, the modal coefficients $f_\mu$ are
\begin{equation}
f_{\mu s}=\psi_s\delta_{\mu,0}.
\end{equation}
To perform the linear stability analysis, we start from the modal equations Eq. (\ref{fnDecompAvg}) linearized around a flat stationary solution $\psi_s$. If we set
\begin{equation}
f_\mu(\tau)=f_{\mu s}+\delta f_\mu(\tau),
\end{equation}
the linearized equations for $f_\mu(\tau)$ read
\begin{multline}
\frac{\partial\delta f_\mu}{\partial\tau}=-\left\{\vphantom{\psi_s^2}(1+i\alpha)\delta f_\mu-ia_\mu\delta f_\mu\right.\\
\left.-i\left[4\delta f_\mu\rho+\delta f_{-\mu}^*\psi_s^2+2\delta_{\mu,0}(\delta f_0^*\psi_s^2+\delta f_0\rho)\right]\right\}\label{linearfn},
\end{multline}
\begin{multline}
\frac{\partial\delta f_{-\mu}^*}{\partial\tau}=-\left\{\vphantom{\psi_s^2}(1+i\alpha)\delta f_{-\mu}^*-ia_\mu\delta f_{-\mu}^*\right.\\
\left.-i\left[4\delta f_{-\mu}^*\rho+\delta f_\mu\psi_s^2+2\delta_{\mu,0}(\delta f_0^*\psi_s^2+\delta f_0\psi_s^{*2})\right]\right\}.\label{linearfnstar}
\end{multline}
A peculiar feature is represented by the terms with the factor $\delta_{\mu,0}$ which appear only in the equations for $\delta f_0$ and $\delta f_0^*$; these terms arise from the last term of Eq. (\ref{fnDecompAvg}) and imply that the case $\mu=0$ must be considered separately from the case $\mu\neq 0$. This feature is not present in the ring cavity. 

If one considers Eqs. (\ref{linearfn}) and (\ref{linearfnstar}) for $\mu=0$, their detailed analysis leads to the usual conclusion: the flat stationary solutions with $\frac{\partial F^2}{\partial\rho}<0$ are unstable. Consequently, if three stationary solutions exist to the FP-LLE at a point $(\alpha,F^2)$ and they are ordered according to magnitude, the middle solution is always unstable.

Next, let us focus on Eqs. (\ref{linearfn}) and (\ref{linearfnstar}) for $\mu\neq 0$. If we set
\begin{equation}
\delta f_\mu(\tau)=e^{\lambda \tau}\delta f_\mu^\prime,\quad \delta f_{-\mu}^*(\tau)=e^{\lambda \tau} \delta f_{-\mu}^{\prime*},
\end{equation}
we obtain a system of linear homogenous equations for $\delta f_\mu^\prime$ and 
$\delta f_{-\mu}^{\prime*}$ that leads to an eigenvalue equation for $\lambda$:
\begin{equation}
\lambda^2+2\lambda+c_o=0,
\end{equation}
with
\begin{equation}
c_o=1+\alpha^2+15\rho^2-8\alpha\rho+2(4\rho-\alpha)a_\mu+a_\mu^2. \label{lambdaeig}
\end{equation}
The solutions of Eq. (\ref{lambdaeig}) are
\begin{equation}
\lambda_\pm=-1\pm\sqrt{1-c_o},
\end{equation}
so that the instability condition $\mathrm{Re\, } \lambda_+>0$ reads
\begin{equation}
\Gamma>1,
\end{equation}
where $\Gamma$ is the gain
\begin{equation}
\Gamma=\sqrt{1-c_o}. \label{Gammadef}
\end{equation}
Hence the flat stationary solution is unstable for $c_o<0$, i.e. for
\begin{equation}
a_{\mu-}<a_\mu<a_{\mu+},\quad a_{\mu\pm}=\alpha-4\rho\pm\sqrt{\rho^2-1},\label{apm}
\end{equation}
or, using Eqs. (\ref{an}) and (\ref{beta}), the pump-referenced optical mode numbers which bound the region of gain $\Gamma>1$ are
\begin{equation}
\mu_\pm=\left\{\frac{2}{\beta}\left[(\alpha-4\rho)\pm\sqrt{\rho^2-1}\right]\right\}^{1/2}, \label{npm}
\end{equation}
where $\mu$ and $a_\mu$ are treated as continuous variables. From Eq. (\ref{npm}) we see that, if $\alpha\leq 4$, the boundary value of $\rho$ at which $\Gamma=1$ (indicating the onset of instability) is $\rho=1$, as in the ring cavity case (see e.g. Refs. \cite{Godey2014}, \cite{Lugiato2015}). Under these conditions the threshold value of $\mu$, determined from $\rho=1$, is
\begin{equation}
\mu_{thr}=\left[\frac{2}{\beta}(\alpha-4)\right]^{1/2}.\label{nthr}
\end{equation}
From Eqs. (\ref{an}), (\ref{beta}), (\ref{lambdaeig}), and (\ref{Gammadef}) we can also obtain the mode number (i.e. the value of $\mu$) for which the gain is maximum, given by 
\begin{equation}
\mu_{max}=\left[\frac{2}{\beta}(\alpha-4\rho)\right]^{1/2}.\label{nmax}
\end{equation}

\begin{figure}[]
	\includegraphics{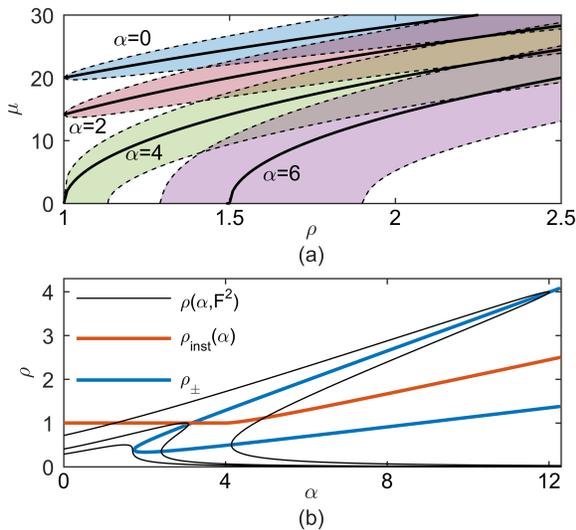} %
	\caption{Exploration of the flat solutions and their stability. (a) Curves $\mu_{max}$ where the gain $\Gamma$ is greatest, for $\beta=-0.02$ and various values of $\alpha$, with shading indicating the region between $\mu_-$ and $\mu_+$ where $\Gamma>1$. Of note is that the region where $\Gamma>1$ does not extend to $\rho=1$ for $\alpha=6$. Regardless of the value of $\beta$, the curve $\mu_{max}$ for $\alpha=4$ passes through the point $(\rho=1,\mu=0)$. (b) Plots of the Kerr-tilted intensity resonance profiles $\rho(\alpha,F^2)$ for three values of $F^2$ (black): $F^2=0.5$ (smallest peak), $1$, and $4$ (largest peak). The orange line shows $\rho_{inst}(\alpha)$, and the blue lines indicate the values $\rho_+(\alpha)$ and $\rho_-(\alpha)$ that bound the region where multiple flat solutions exist. Resonance curves for $F^2>1$ are qualitatively similar to the curve for $F^2=4$. \label{Fig3}}
\end{figure}

Figure \ref{Fig3}a shows the curves $\mu_+$, $\mu_-$, and $\mu_{max}$ as functions of $\rho$ for the fixed value of $\beta=-0.02$ and various values of $\alpha$. From this figure, it is apparent that when $\alpha>4$ (and $\beta<0$, $|\beta|\ll1$), instability exists above the value of $\rho$ at which $\mu_- (\rho)$ vanishes; this value depends on $\alpha$ according to Eq. (\ref{npm}). Therefore, we denote by $\rho_{inst}(\alpha)$ the curve defined by $\rho=1$ below $\alpha=4$  and $\mu_- (\rho)=0$ above it, leading to $\rho_{inst}(\alpha)=(4\alpha-\sqrt{\alpha^2-15})/15$ for $\alpha>4$. 

The value of $\rho_{inst}(\alpha)$ is shown in Fig. \ref{Fig3}b, together with the functions $\rho_\pm (\alpha)$ given by Eq. (\ref{derF2zero}), which define the limits of the upper and the lower branch of stationary solutions. We also plot curves illustrating the stationary solutions $\rho(\alpha,F^2 )$ for three values of $F^2$. A stationary solution $\psi_s$ ($|\psi_s |^2=\rho$) is unstable if it lies above the line $\rho_{inst}(\alpha)$. Further, the solution is unstable if $\partial\rho/\partial F^2 <0$, i.e. if it is the middle of three values of $\rho(\alpha,F^2)$, regardless of its location relative to $\rho_{inst} (\alpha)$. Therefore, the curve $\rho_{inst} (\alpha)$ does not correspond to a stability boundary when it intersects the middle branch of the stationary curve, because this curve is already unstable. This is the case for intersections of the stationary curves with $\rho_{inst} (\alpha)$ when $\alpha>\alpha_1$, where $\alpha_1\approx3.17$ is the value at which $\rho_+ (\alpha)=1$, given by $(2\alpha_1+\sqrt{\alpha_1^2-3})/9=1$ and occurring at the intersection of the dashed red $\rho_{inst}(\alpha)$ curve with the lower blue curve in Fig. \ref{Fig2}b. 

From an experimental standpoint, we are concerned with identifying the values $\alpha$ at which the flat solution becomes unstable and an extended pattern will be formed as $\alpha$ is varied. From Figs. \ref{Fig2}b and \ref{Fig3}b we can see that, if we increase the value of $\alpha$ from some negative initial value (this corresponds to decreasing the laser frequency from blue detuning), for $F^2>1$ the flat solution becomes unstable when $\rho=1$, while for $F^2<1$ it is always stable. On the other hand, if we decrease the value of $\alpha$ from some large positive initial value (this corresponds to increasing the laser frequency from red detuning), the system will follow the lower branch, which is always stable, until this branch vanishes, whereupon a jump to the unique and unstable (if $F^2>1$) flat solution results in the formation of an extended pattern. In the $\alpha-F^2$ plane shown in Fig. \ref{Fig2}b, the disappearance of the lower branch (see Fig. \ref{Fig3}b) corresponds to the upper blue line; in Fig.~\ref{Fig3}b this disappearance corresponds to the intersection of the lower branch of the black curves with the lower blue line.

In the case of the ring cavity the factor $4$ must be replaced with $2$ in Eqs. (\ref{apm})-(\ref{nmax}).

\subsection{General connection between stationary patterns of the LLE in the ring and FP geometries}

Let us now focus on the stationary patterns, in which $\psi$ depends on $\theta$ but not on $\tau$. Equation (\ref{FPLLE}) reduces to
\begin{equation}
F=\left[1+i(\alpha-2\left<|\psi|^2\right>)\right]\psi+i\frac{\beta}{2}\frac{\partial^2\psi}{\partial\theta^2}-i|\psi|^2\psi,
\end{equation}
which shows that a stationary pattern for the FP cavity for the parameter $\alpha$ coincides with a stationary pattern for the ring cavity when $\alpha$ is replaced with $\alpha^\prime$, given by
\begin{equation}
\alpha^\prime=\alpha-2\left<|\psi|^2\right>. \label{alphaprime}
\end{equation}

Note that $\alpha^\prime$ depends on the shape of the pattern. One caveat is that this connection does not formally extend to the stability of stationary patterns, so to establish stability the analysis must be performed in each case separately.

\begin{figure}[]
	\includegraphics{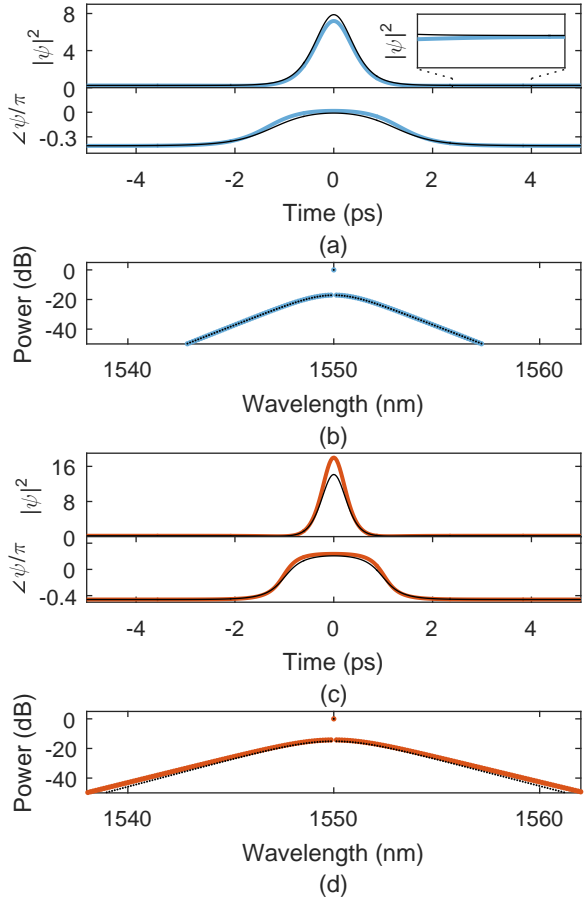} %
	\caption{Soliton solutions to the FP-LLE. Analytical approximations are shown in black; numerically calculated solutions are shown in color. Here $\beta=-0.02$ and the resonator FSR is $16.5$ GHz. (a, b) $F^2=3$, $\alpha=4.37$, time domain curves in (a) with inset showing the deviation between analytic approximation and numerical solution in the level of the c.w. background near the pulse, optical spectrum in (b). (c,~d) $F^2=12$, $\alpha=8.68$. \label{Fig4}}
\end{figure}

\section{Analytical approximation of solitons\label{sec5}}

There are two fundamental types of stationary patterns: Turing patterns and solitons. Turing patterns are periodically modulated solutions with a number of maxima and minima throughout the domain, while solitons are single-peaked. When the stationary curve is single-valued, Turing patterns arise when the instability threshold is crossed. Analytic approximations for these patterns are possible near threshold \cite{Lugiato1987,Lugiato1987a} or in the small damping limit \cite{Renninger2016}. We focus our immediate attention on analytic approximations for solitons, and discuss Turing patterns below. Because of the correspondence between stationary patterns of the ring LLE and stationary patterns of the FP-LLE, discussed in the previous section and summarized by Eq. (\ref{alphaprime}), we begin by recalling the approximation to solitons for the ring LLE. Stationary solutions to the ring LLE satisfy Eq. (\ref{FPLLE}) without the derivative with respect to time and without the integral term. The expression \cite{Herr2014wArxiv} 

\begin{equation}
\psi_{sol}(\theta)=\psi_s+\sqrt{2\alpha}e^{i\phi_o}\sech{\left(\sqrt{\frac{2\alpha}{-\beta}}\theta\right)}, \label{psisolCW}
\end{equation}
approximates the stationary soliton solution to the ring LLE, including a constant background corresponding to the stable flat stationary solution in the lower branch of the stationary curve. Here, 
\begin{equation}
\phi_o=\cos^{-1}(\sqrt{8\alpha}/\pi F),
\end{equation}
and $\psi_s$ denotes the unique flat stationary solution when $\alpha<\sqrt{3}$ and the flat stationary solution in the lower branch of the stationary curve when $\alpha>\sqrt{3}$. The relevant stationary equation is for the ring LLE, and is identical to Eq. (\ref{statF}) with $3$ replaced with $1$:
\begin{equation}
F=\left[1+i(\alpha-\rho)\right]\psi_s,
\end{equation}
where $\rho=|\psi_s|^2$.
Two important characteristics of the function in Eq. (\ref{psisolCW}) are worth noting, because they remain generally true for soliton solutions of the ring LLE and the FP-LLE: The amplitude of the soliton increases with increased detuning $\alpha$, and the temporal width of the soliton decreases as $\alpha$ is increased or $\beta$ is decreased. We note that an approximation without the background, which is an exact solution to the ring LLE with a particular form of a spatially-varying pumping term $F^2(\theta)$, is presented in Ref. \cite{Coen2013}.

Let us now turn to the case of the FP cavity. On the basis of the general connection between stationary patterns in the ring and in the FP case and using Eq. (\ref{psisolCW}), we can write that the approximate analytic expression of the soliton is:

\begin{equation}
\psi_{sol}(\theta)=\psi_s^\prime+\sqrt{2\alpha^\prime}e^{i\phi_o^\prime}\sech{\left(\sqrt{\frac{2\alpha^\prime}{-\beta}}\theta\right)}, \label{FPpsisolCW}
\end{equation}
where by integrating Eq. (\ref{FPpsisolCW}) and using Eq. (\ref{alphaprime}) we obtain an equation for $\alpha^{\prime}$:
\begin{multline}
\alpha^\prime=\alpha-2\rho^\prime-\frac{2}{\pi}\sqrt{-2\alpha^\prime\beta}\tanh\left(\pi\sqrt{\frac{2\alpha^\prime}{-\beta}}\right)\\
-\frac{8\sqrt{-\beta\rho^\prime}}{\pi}\cos(\phi^\prime-\phi_o^\prime)\tan^{-1}{\tanh{\left(\pi\sqrt{\frac{\alpha^\prime}{-2\beta}}\right)}}.
\end{multline}

Here quantities have been primed to indicate that they are defined according to the ring LLE at the point $(\alpha^\prime,F^2)$ in the parameter space plane, thus: 
\begin{align}
F^2&=\rho^\prime\left(1+(\alpha^\prime-\rho^\prime)^2\right)\label{rhoprime},\\
\psi_s^\prime&=\frac{F}{1+i(\alpha^\prime-\rho^\prime)},\\
\phi^\prime&=\arg(\psi_s^\prime)=\tan^{-1}(\rho^\prime-\alpha^\prime),\\
\phi_o^\prime&=\cos^{-1}\frac{\sqrt{8\alpha^\prime}}{\pi F}\label{phioprime}.
\end{align}

In Fig. \ref{Fig4} we present plots of the analytical approximation to the soliton solution of the FP-LLE as described by Eqs. (\ref{FPpsisolCW}-\ref{phioprime}). For comparison, we also present numerically calculated steady-state soliton solutions of the FP-LLE. These simulations, and the numerical investigations of the FP-LLE presented in the following sections, are performed using a fourth-order Runge-Kutta interaction picture method \cite{Hult2007} with an adaptive step size \cite{Heidt2009}.

In addition to single solitons, the FP-LLE supports ensembles of multiple co-propagating solitons as stationary solutions, with an analytical approximation to these ensembles possible as 

\begin{multline}
\psi_{ens.}(\theta)=\psi_s^\prime\\+\sqrt{2\alpha^\prime}e^{i\phi_o^\prime}\sum_{j}\sech{\left(\sqrt{\frac{2\alpha^\prime}{-\beta}}(\theta-\theta_j)\right)}.\label{solensemble}
\end{multline}

Such an ensemble may or may not be stable, depending on the separation between the locations of the solitons $\{\theta_j\}$ and the temporal width of the solitons determined by $\alpha^\prime$ and $\beta$. Each soliton in the ensemble contributes to the average intensity $\left<|\psi|^2\right>$, and so a different equation for $\alpha^\prime$ must be derived by integrating Eq. (\ref{solensemble}) over the domain.

\section{Turing patterns\label{sec6}}
\begin{figure}[]
	\includegraphics{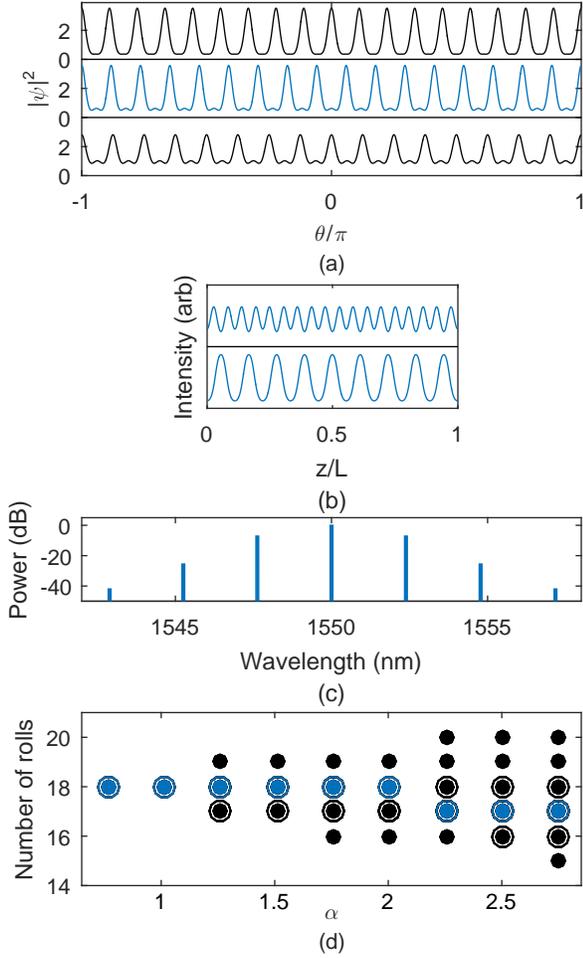} %
	\caption{Simulations of Turing patterns in the FP-LLE. (a)~Three Turing patterns, with $18$, $17$, and $16$ rolls, all stable against perturbations at the point $(\alpha=2.5,F^2=6)$. The Turing pattern with $17$ rolls, plotted in blue, arises most frequently from vacuum fluctuations. (b) Plots of $|F_F |^2+|F_B |^2$ in the physical domain $0<z<L$ at two different times for the $17$-roll Turing pattern plotted in (a), demonstrating how a stationary solution of the FP-LLE relates to a time-varying intensity pattern. (c) Optical spectrum of the Turing pattern plotted in blue from (a) and (b), assuming a cavity with $16.5$~GHz FSR. (d) Summary of multi-stabilility of Turing patterns as revealed by simulations for $F^2=6$. Data points indicate Turing patterns that can be excited from appropriate initial conditions. Data points enclosed by a circle indicate Turing patterns stable against perturbations, and blue data points indicate Turing patterns that arise from vacuum fluctuations. \label{Fig5}}
\end{figure}

In this section we discuss the formation and behavior of Turing patterns, with results summarized in Figure \ref{Fig5}. Turing patterns have great relevance for experiment because they can be generated with a blue-detuned pump laser, avoiding thermal instability \cite{Carmon2004}. Moreover, they are the first non-c.w. phenomenon generated in a decreasing scan of the pump-laser frequency across a cavity resonance, a technique that has been used to explore Kerr nonlinear optics and generate solitons in ring resonators. 

As in the ring cavity, Turing patterns in the FP cavity are generated spontaneously through the amplification of vacuum fluctuations and other noise. This amplification is caused by modulation instability of the flat solution when its amplitude is above the instability threshold, which as discussed above occurs when $\rho>1$ if $\alpha<4$, and when $\rho$ is above the value determined by $\mu_- (\alpha)=0$ otherwise. In the field quantity $\psi$ of the FP-LLE, a Turing pattern consists of a periodically modulated waveform with multiple minima and maxima in $|\psi|^2$ over the domain of length $2L$. Corresponding to the $n$-fold decreased period (relative to the round-trip time) of an $n$-roll Turing pattern’s modulated waveform in the time-domain, the optical spectrum of a Turing pattern consists of modes spaced by $n$ resonator FSR. Fig. \ref{Fig5}a shows plots of $|\psi|^2$ for several Turing patterns that are stable at the point $(\alpha=2.5,F^2=6)$, and Fig.~\ref{Fig5}b shows plots of the representative physical quantity $|F_F |^2  +|F_B |^2$, which is proportional to the intensity in the resonator averaged over fast temporal and spatial oscillations associated with the optical frequency, for one of these. A difference between Turing patterns in ring resonators and Turing patterns in Fabry-Perot resonators is that the intensity profile is constant, up to rotation at the group velocity, in the ring resonator. In the Fabry-Perot resonator, however, the intensity profile evolves with time in a more complex way due to the summation of the intensities of counter-propagating waves; this phenomenon is demonstrated by the two intensity profiles depicted in Fig. \ref{Fig5}b. Fig. \ref{Fig5}c shows an optical spectrum corresponding to this Turing pattern, assuming a cavity FSR of~$16.5$ GHz.

The multiplicity (or number of rolls) $n$ of a Turing pattern excited from broadband noise is determined by the spectrum of the modulation instability gain and the presence of noise to seed the formation of the pattern, and is close to the number $\mu_{max}=\left[\frac{2}{\beta} (\alpha-4\rho)\right]^{1/2}$ described above. This process is not deterministic, but from broadband noise the distribution of roll numbers is very narrow\textemdash running $1000$ trials of Turing pattern generation from white noise at the point $\alpha\approx2.75$ and $F^2=6$ yielded $997$ Turing patterns with $17$ rolls, one Turing pattern with $16$ rolls, and two Turing patterns with $18$ rolls.

Despite the narrow distribution of Turing pattern multiplicity $n$ generated from broadband noise, multiple Turing patterns with different $n$ values can exist stably at the same point in the $\alpha-F^2$ plane. As $\rho$ increases due to increased pump-laser power or wavelength, the range of stable $n$ values increases in accordance with the increase in the difference $\mu_+-\mu_-$ (see Eq. (\ref{npm})). In Fig.~\ref{Fig5}d we plot, for $F^2=6$ and for various values of $\alpha$, the multiplicity $n$ of Turing patterns we have generated in simulations.

\section{Nonstationary solutions of the FP-LLE\label{sec7}}
\begin{figure}[]
	\includegraphics{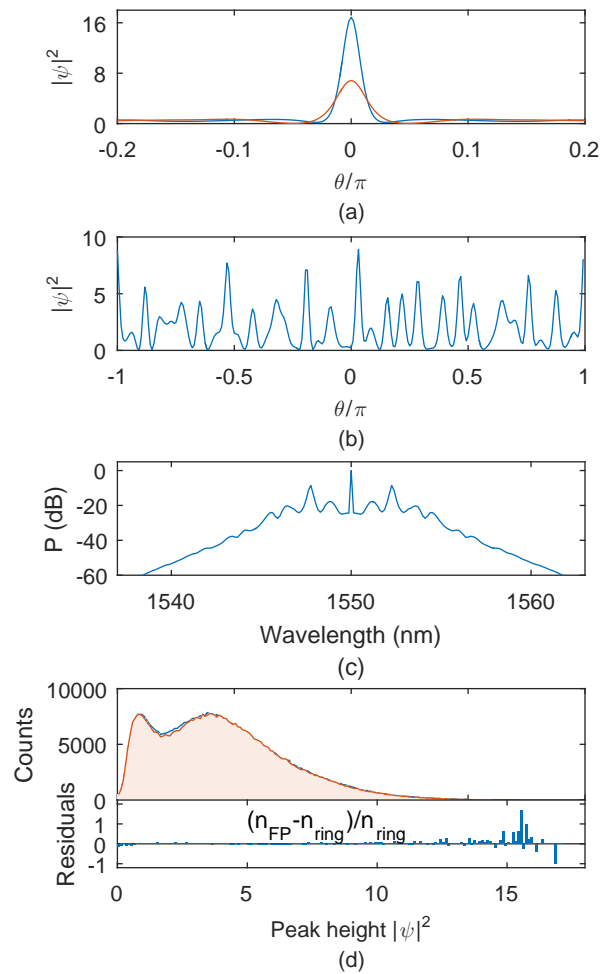} %
	\caption{Numerical investigations of nonstationary solutions to the FP-LLE. (a) Maximum (blue) and minumum (orange) amplitudes of an oscillating breather soliton at the point $(\alpha=5.6,F^2=8)$ for $\beta=-0.02$. (b) A snapshot of the time-varying, aperiodic intensity profile of spatiotemporal chaos at the point $(\alpha=5.3,F^2=8)$ for $\beta=-0.02$. (c) Time-averaged optical spectrum of spatiotemporal chaos under the conditions given in (b) for a cavity with $16.5$ GHz FSR. (d) Top: A histogram of local maxima values of spatiotemporal chaos for the FP LLE at the point $(\alpha=5.3,F^2=8)$, $\beta=-0.02$ (blue), and at the corresponding point $(\alpha=1.1,F^2=8)$ for the ring LLE (orange), recorded over simulations with the same duration. Bottom: Fractional difference between the two histograms.\label{Fig6}}
\end{figure}

The stationary solutions of the FP-LLE are the flat solutions, solitons and soliton ensembles, and Turing patterns. Besides these, the FP-LLE exhibits the same nonstationary solutions as the LLE for the ring cavity: spatiotemporal chaos and breather solitons. These solutions can be investigated numerically, and some results of these investigations are shown in Figure \ref{Fig6}. Spatiotemporal chaos consists of many fluctuating and colliding pulses that fill the cavity. Generally, chaos lies in a region of the $\alpha-F^2$ plane which is reached by increasing $\alpha$ or $F^2$ from a point where Turing patterns exist \cite{Godey2014}, provided $F^2$ is above some critical threshold, and the fluctuations of the chaos become more severe as the pump power $F^2$ is increased. The transition from Turing patterns to chaos is not well defined, but begins with a kind of period-doubling of the Turing pattern in the time-domain in which the amplitudes of the Turing pattern rolls begin to oscillate, with adjacent rolls oscillating out of phase. Following this, pulses begin to exhibit lateral motion and collisions, and the number of maxima and minima in the cavity then varies \cite{Coillet2014}. Breather solitons are pulses whose amplitudes oscillate periodically, and they are found near the lower bound in $\alpha$ of the region where solitons can exist (discussed extensively below). The properties of these phenomena are similar to their ring LLE counterparts.

An interesting question is whether the dynamics of spatiotemporal chaos under the FP-LLE differ significantly from the dynamics under the ring LLE as a result of the fluctuations in the value of $\alpha^\prime=\alpha-2\left<|\psi|^2\right>$ attendant to the fluctuations in the average intensity $\left<|\psi|^2\right>$. As a preliminary investigation, we perform simulations of spatiotemporal chaos under the FP-LLE at the point $(\alpha=5.3,F^2=8)$, and then perform simulations under the ring LLE at the point $(\overline{\alpha^\prime}=\alpha-2\overline{\left<|\psi|^2\right>}=1.1,F^2=8)$, where $\bar{g}$ denotes time-averaging. The histogram of the height of local maxima in $|\psi|^2$ shown in Fig. \ref{Fig6}d suggests little difference in the behavior of chaos between the two equations, but more extensive investigations could yield more interesting results.

\section{Practical implications of the nonlinear integral term\label{sec8}}
\begin{figure}[]
	\includegraphics{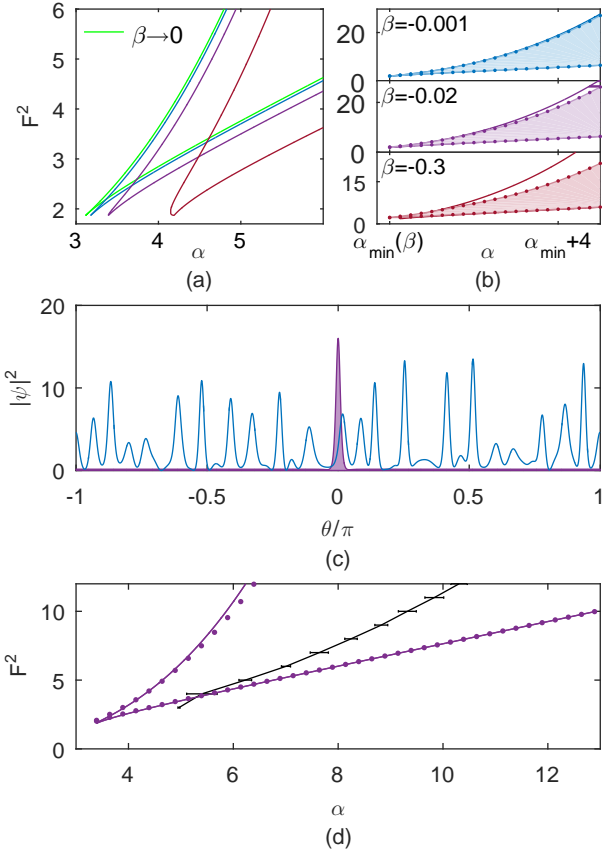} %
	\caption{Effects of nonlinear integral term in the FP-LLE, Eq. (\ref{FPLLE}). (a) Approximate existence bounds of single solitons in the zero-dispersion limit (green) and for $\beta=-0.001$ (blue), $\beta=-0.02$ (purple), and $\beta=-0.3$ (red), calculated using the analytical approximation to the single soliton. (b) Comparisons between the approximate existence bounds and the bounds as determined numerically. Solid points indicate the maximum and minimum values of $F^2$ at which solitons have been simulated for a given value of $\alpha$. (c) Simulated spatiotemporal chaos (blue) and single soliton solution (purple), either of which can exist at the point $(\alpha=8,F^2=8)$. The amplitude of the soliton is larger than the characteristic amplitude of the features in the chaos because the effective detuning $\alpha^\prime$ is larger for the soliton. (d) Analytical and numerical soliton existence limits (purple) for $\beta=-0.02$ from panel (a) and the upper bound in $\alpha$ for the existence of spatiotemporal chaos/Turing patterns (black), estimated as described in the text. \label{Fig7}}
\end{figure}

In this section we discuss implications of the differences between the FP-LLE and the ring LLE\textemdash namely, the additional nonlinear integral term representing modulation by twice the average intensity\textemdash for the types of experiments that have been conducted in Kerr ring resonators. Many of the promising potential applications of c.w.-pumped Kerr ring and FP resonators rely on the generation of single solitons, so we focus here on two issues: the effect of the nonlinear integral term on the existence range of single solitons, and its effect on the experimental generation of single solitons via scans of the pump-laser frequency. These results are summarized in Fig. \ref{Fig7}. 

\subsection{Existence range of single solitons}
An important consequence of the additional nonlinear term $2i\psi\left<|\psi|^2\right>$ in the FP-LLE is that the range of parameters over which single solitons exist acquires a dependence on the dispersion parameter $\beta$, through the effect of dispersion on pulse energy.  This is in contrast to the situation for the ring LLE, where the existence range is independent of $\beta$. For the FP-LLE the existence range also depends on the number of co-propagating pulses, but we will not discuss this at length here. 

As is well-known for the ring LLE, solitons can exist only with a red-detuned pump laser $\alpha>0$ so that the phase rotation coming from the detuning term $\alpha$ in the LLE can be balanced by the phase shift from the nonlinear terms. The minimum value of detuning $\alpha$ at which solitons exist as a function of $F^2$ is determined by the existence of a stable flat solution to the LLE which can form the c.w. background for the soliton. The maximum value of detuning for which solitons can exist is determined by $\alpha^\prime=\alpha-2\left<|\psi|^2\right>$ according to $\alpha_{max}^\prime(F^2)=\pi^2 F^2/8$, which approximately gives the maximum detuning for solitons in the ring LLE \cite{Herr2014wArxiv}. 

For the FP-LLE, a stable flat solution exists to the right of the line $F_+^2 (\alpha)$ in the $\alpha-F^2$ plane which bounds from above the region of multiple flat solutions, shown by the upper blue line in Fig. \ref{Fig2}b. Explicitly (see Eqs. (\ref{statF2}) and (\ref{derF2zero}) and the accompanying discussion), this curve in the $\alpha-F^2$ plane is given by
\begin{align}
F_+^2(\alpha)&=F^2\left(\rho_-(\alpha),\alpha\right),\label{alphamin1}\\ 
F^2(\rho,\alpha)&=\rho\left(1+(\alpha-3\rho)^2\right),\label{alphamin2}\\
\rho_-(\alpha)&=\left(2\alpha-\sqrt{\alpha^2-3}\right)/9, \label{alphamin3}
\end{align}
so that the minimum detuning for solitons $\alpha_{min}(F^2)$ in the limit $\beta\rightarrow0^-$ (leading to zero soliton energy) is determined by inverting Eqs. (\ref{alphamin1})-(\ref{alphamin3}) to solve for $\alpha$ as a function of $F^2$. In the same limit of zero soliton energy, the maximum value of detuning for the FP-LLE at fixed $F^2$ is approximately $\alpha_{max}(F^2)=\alpha_{max}^\prime+2\rho_{min}^\prime\left(\alpha_{max}^\prime(F^2),F^2\right)$, where $\rho_{min}^\prime(\alpha^\prime,F^2)$ is the smallest solution to $F^2=\rho^\prime\left[1+(\alpha^\prime-\rho^\prime)^2\right]$ (here, as before, primed values are solutions of the appropriate equations for the ring LLE). These boundaries are plotted in Fig.~\ref{Fig7}a.

For finite dispersion and soliton energy, numerical simulations show that the curves bounding the region of soliton existence are shifted as the energy of the soliton changes with dispersion. An intuitive way to understand the shift of the left boundary $\alpha_{min}$ is to note that the introduction of a soliton with finite energy onto a stable flat solution near and just right of the boundary $\alpha_{min}$ leads to a decrease in the effective detuning $\alpha_{eff}$ under which the flat solution evolves (due to the nonlinear integral term), and if this decrease is large enough it can lead to instability.

A preliminary approximation to the dispersion-dependent boundary curves can be obtained using the analytical approximation to the soliton solution given by Eqs. (\ref{FPpsisolCW}-\ref{phioprime}). For fixed $\beta$ and $F^2$, we calculate the values of $\alpha$ at which the background $\rho^\prime$ of the soliton (given by Eq. (\ref{rhoprime})) is the same as the background along the zero-dispersion boundary curves; that is, for the dispersion-shifted left boundary we use the requirement $\rho^\prime=\rho_-$, and for the dispersion-shifted right boundary we use the requirement $\rho^\prime=\rho_{min}^\prime (\alpha_{max}^\prime (F^2),F^2)$. We compare the resulting curves with a numerical determination of the boundary curves for three finite values of dispersion; the results are plotted in Figs. \ref{Fig7}a and b. The analytical approximation is accurate for low $F^2$ and small dispersion, but becomes less accurate as these quantities increase. This is because breather solitons are found near $\alpha_{min}$ for larger values of $F^2$. Breather solitons are accompanied by traveling waves that propagate away from the soliton and diminish in amplitude as they do so, and their range increases with the dispersion. For larger values of dispersion these waves fill the cavity, and in this case the flat background whose stability forms the basis for approximating the dispersion-dependent boundary curves is actually not present. 

The lines $\alpha_{min} (F^2)$ and $\alpha_{max} (F^2)$ intersect at $F^2=F_I^2\approx1.87$. Below this value of the pump power solitons do not exist for the FP-LLE, and this can be seen as follows: The value of $\rho_{min}^\prime$ describing the amplitude of the soliton background along the line of maximum detuning $\alpha_{max}^\prime$ for the ring LLE is in general also a flat solution $\rho$ of the FP-LLE at the corresponding point $\alpha_{max}=\alpha_{max}^\prime+2\rho_{min}^\prime (\alpha_{max}^\prime,F^2)$; this is due to the general correspondence between stationary patterns discussed in Sec. \ref{sec4}c. However, when $F^2<F_I^2\approx1.87$, the flat solution $\rho_{min}^\prime$ to the ring LLE is not the smallest flat solution to the FP-LLE; instead, it is the middle of three, and is therefore unstable. Therefore, when $F^2<F_I^2$ the line $\alpha_{max} (F^2)$ as defined above does not represent the right boundary of soliton existence for the FP-LLE. In fact, below this point, for all values of $\alpha$ where a stable flat solution to the FP-LLE $\rho_{min}$   exists, $\alpha-2\rho_{min} (\alpha,F^2 )>\alpha_{max}^\prime$, preventing the existence of solitons.

\subsection{Generation of single solitons through laser frequency scans}

A second important consequence of the additional nonlinear term is an increase in the range of $\alpha$ values, for a given value of $F^2$, at which the state of $\psi$ can be either an extended pattern (spatiotemporal chaos or Turing pattern) or a soliton/soliton ensemble. This is because the extended patterns fill the domain and, because of their higher average intensity, experience a greater nonlinear shift than lower duty-cycle single solitons or soliton ensembles due to the nonlinear integral term. In Fig. \ref{Fig7}c we plot simulations of spatiotemporal chaos and a single soliton, both of which can be obtained at the point $(\alpha=8,F^2=8)$, along with a stable flat solution, depending on the initial conditions. 

A practical consideration is the impact of the nonlinear integral term on the generation of single solitons via the well-established method of scanning the laser across the pumped resonance with decreasing frequency (increasing $\alpha$)\cite{Herr2014wArxiv}. Because this method relies on the excitation of an extended pattern (chaos or Turing pattern) to provide initial conditions out of which solitons condense as $\alpha$ is increased, it is important that the maximum detuning (the value of $\alpha$ where $\alpha^\prime=\alpha_{max}^\prime=\pi^2 F^2/8)$ for single solitons is larger than the $\alpha$ value at which an extended pattern will transition to a soliton ensemble. Otherwise, the generation of single solitons using this method will be difficult or impossible. To investigate this, we numerically perform slow scans across the resonance to identify where the transition from extended patterns to independent solitons occurs. These scans are conducted slowly to approximate adiabaticity: $d\alpha/d\tau=2.5\times10^{-4}$. We perform $10$ scans across the resonance at each integer value of $F^2$ from $3$ to $12$ with $\beta=-0.02$, and we identify the transition from extended pattern to independent solitons by inspection of several quantities as $\alpha$ is varied: the set of local maxima and minima of $|\psi|^2$ (see \cite{Coillet2014}), the distance between local maxima, and the number of local maxima above $|\psi|^2=1$. In Fig. \ref{Fig7}d we plot the line representing the upper boundary in $\alpha$ of extended patterns obtained in the scans across the resonance. Error bars represent the standard deviation of the values $\alpha$ at which the transition is observed, with this spread in the values arising due to the chaotic fluctuations in the total intracavity power and therefore also in the size of the nonlinear integral term. These results indicate that the region over which single solitons exist and extended patterns do not is narrow for small pump powers $F^2$, and widens as $F^2$ is increased. Without performing experiments, it is impossible to precisely quantify the limitations imposed by this observation, but we expect this finding to be useful in refining schemes for single-soliton generation. These challenges associated with the necessary transition from high duty-cycle extended patterns to low duty-cycle solitons are alleviated by pulsed pumping, as demonstrated in Ref. \cite{Obrzud2017}.

For completeness, we note that after the transition to solitons, we observe in a small number of scans the onset of harmonic modelocking, by which we mean a slow (i.e. over hundreds or thousands of photon lifetimes) convergence of the soliton ensemble to uniform spacing. Because harmonic modelocking eliminates soliton collisions, which are the mechanism by which single solitons can be obtained from a multi-soliton ensemble, it is unclear whether single solitons can be generated under these conditions. Harmonic modelocking is clearly observed for all the scans for $F^2=3$, four of the scans for $F^2=7$, and one scan for $F^2=9$. While we have not investigated the phenomenon in depth, we speculate that harmonic modelocking occurs when a soliton ensemble exhibits a suitable initial distribution and an appropriate density of solitons, which is related to the pulse width and therefore to $\alpha$ and $\beta$.

\section{Conclusions\label{sec9}}

In this article we have presented the analogue of the spatiotemporal Lugiato-Lefever equation applicable to a Fabry-Perot cavity with the Kerr nonlinearity. This equation is derived from the appropriate Maxwell-Bloch equations and allows determination of the forward- and backward-propagating field components in the cavity. We have presented this equation using notation that is standard in the field of microresonator-based frequency combs and that makes clear the important difference between the two geometries, which is the existence of an additional nonlinear integral term representing modulation by twice the average intensity. This term leads to subtle but important differences in the dynamics and the stationary states exhibited in the two geometries. We expect that our preliminary investigation of these differences will facilitate future experiments using the Fabry-Perot geometry. 

Importantly, the states that are stationary for the ring LLE (solitons and Turing patterns) are also stationary for the FP-LLE, up to a shift in the cavity detuning parameter $\alpha$. As discussed in Sec. \ref{sec8}, this shift has implications for experimental generation of cavity solitons. Besides this, we have described our observations in simulation of multi-stability of Turing patterns under the FP-LLE, and also investigated the nonstationary solutions, spatiotemporal chaos and breather solitons. These states appear similar to their counterparts under the ring LLE, but further investigation of their properties may reveal interesting differences due to the fact that these states evolve with a fluctuating effective detuning parameter $\alpha^\prime$.

The Fabry-Perot geometry represents an exciting new direction for frequency comb generation in passive Kerr resonators, as indicated also by the work of Obrzud \textit{et al.} \cite{Obrzud2017}. This geometry presents a different set of cavity properties (e.g. wavelength-dependent mirror-coating reflectivity and group-delay dispersion) that can be manipulated to control the properties of the frequency comb, and thus has the potential to expand the range of applications available to this emerging technology.

\bibliography{ms}

\begin{acknowledgments}
We thank Daniel Hickstein and Liron Stern for comments on the manuscript. This work was supported by NIST, NASA, and the NSF under grant no. DGE 1144083. This work is a contribution of the US government and is not subject to copyright in the United States.
\end{acknowledgments}

\end{document}